\documentclass[twocolumn,
amsmath,amssymb,superscriptaddress,aps,
reprint,
pra,
floatfix
]{revtex4}

\usepackage{float}
\usepackage[T1]{fontenc}
\usepackage{graphicx}

\usepackage[caption=false]{subfig} 
\usepackage{ragged2e} 
\DeclareCaptionJustification{justified}{\justifying}

\usepackage{dcolumn}
\usepackage{bm}
\usepackage{academicons}
\usepackage{hyperref}
\hypersetup{colorlinks}
\usepackage{pifont}
\usepackage{natbib}
\usepackage{titlesec}
\usepackage{soul}
\usepackage{blindtext}
\usepackage{xcolor}
\usepackage{tikz}
\usepackage{svg}
\usepackage{subfig}
\usepackage{lipsum} 

\newcommand{\orcidicon}[1]{\href{https://orcid.org/#1}{\includegraphics[height=\fontcharht\font`\B]{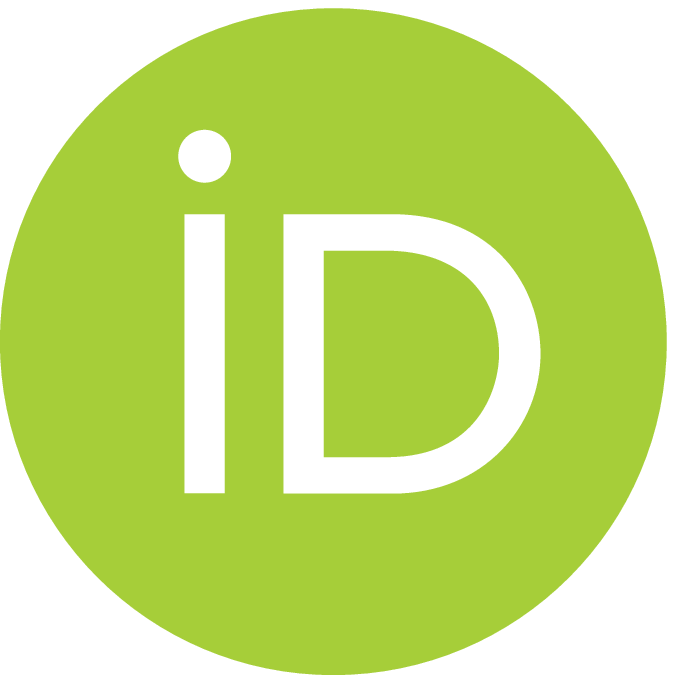}}}

\hypersetup{
colorlinks = true,
urlcolor   = blue,
linkcolor  = red,
citecolor  = blue
}

\begin{document}

\preprint{APS/PRB}
\title{Influence of the Magnetic Sub-Lattices in the Double Perovskite Compound LaCaNiReO$_6$}

\author{Konstantinos~Papadopoulos \orcidicon{0000-0002-4439-4721} }
\affiliation{Department of Physics, Chalmers University of Technology, SE-41296 G\"oteborg, Sweden}
\author{Ola~Kenji~Forslund}
\affiliation{Department of Applied Physics, KTH Royal Institute of Technology, SE-106 91 Stockholm, Sweden}
\author{Elisabetta~Nocerino}
\affiliation{Department of Applied Physics, KTH Royal Institute of Technology, SE-106 91 Stockholm, Sweden}
\author{Fredrik~O.L.~Johansson
\orcidicon{0000-0002-6471-1093}}
\affiliation{Division of Molecular and Condensed Matter Physics, Uppsala University, SE-752 37 Uppsala, Sweden}
\affiliation{Division of Applied Physical Chemistry, KTH Royal Institute of Technology, SE-100 44 Stockholm, Sweden}
\affiliation{Sorbonne Universite, UMR CNRS 7588, Institut des Nanosciences de Paris, F-75005 Paris, France}
\author{Gediminas Simutis}
\affiliation{Laboratory  for  Neutron  and  Muon  Instrumentation, Paul  Scherrer  Institut,  CH-5232  Villigen  PSI,  Switzerland}
\affiliation{Department of Physics, Chalmers University of Technology, SE-41296 G\"oteborg, Sweden}
\author{Nami~Matsubara}
\affiliation{Department of Applied Physics, KTH Royal Institute of Technology, SE-106 91 Stockholm, Sweden}
\author{Gerald~Morris}
\author{Bassam~Hitti}
\author{Donald~Arseneau}
\affiliation{TRIUMF, 4004 Wesbrook Mall, Vancouver, BC, V6T 2A3, Canada}
\author{Peter~Svedlindh}
\affiliation{Uppsala University, Department of Materials Science and Engineering, Uppsala University, 751 03, Uppsala, Sweden}
\author{Marisa~Medarde}
\affiliation{Laboratory for Multiscale materials eXperiments, Paul  Scherrer  Institut,  CH-5232  Villigen  PSI,  Switzerland}
\author{Daniel~Andreica}
\affiliation{Faculty of Physics, Babes-Bolyai University, 400084 Cluj-Napoca, Romania}
\author{Vladimir~Pomjakushin}
\affiliation{Laboratory for Neutron Scattering and Imaging, Paul Scherrer Institute, CH-5232, Villigen, PSI, Switzerland}
\author{Lars B\"orjesson}
\affiliation{Department of Physics, Chalmers University of Technology, SE-41296 G\"oteborg, Sweden}
\author{Jun~Sugiyama \orcidicon{0000-0002-0916-5333}}
\affiliation{Neutron Science and Technology Center, 
Comprehensive Research Organization for Science and Society (CROSS), Tokai, Ibaraki 319-1106, Japan}
\author{Martin~M\aa nsson}
\affiliation{Department of Applied Physics, KTH Royal Institute of Technology, SE-106 91 Stockholm, Sweden}
\author{Yasmine~Sassa
\orcidicon{0000-0003-1416-5642}}
\email{yasmine.sassa@chalmers.se}
\affiliation{Department of Physics, Chalmers University of Technology, SE-41296 G\"oteborg, Sweden}

\date{\today}

\begin{abstract}
\vspace{6mm}
The magnetism of double perovskites is a complex phenomenon, determined from intra- or interatomic magnetic moment interactions, and strongly influenced by geometry. We take advantage of the complementary length and time scales of the muon spin rotation, relaxation and resonance ($\mu^{+}$SR) microscopic technique and bulk AC/DC magnetic susceptibility measurements to study the magnetic phases of the LaCaNiReO$_{6}$ double perovskite. As a result we are able to discern and report a newly found dynamic phase transition and the formation of magnetic domains below and above the known magnetic transition of this compound at $T_{\rm N}=103$~K. $\mu^{+}$SR, serving as a local probe at crystallographic interstitial sites, reveals a transition from a metastable ferrimagnetic ordering below $T=103$~K to a stable one below $T=30$~K. The fast and slow collective dynamic state of this system are investigated. Between $103~{\rm K}<T<230~{\rm K}$, the following two magnetic environments appear, a dense spin region and a static-dilute spin region. The paramagnetic state is obtained only above $T>270$~K. An evolution of the interaction between Ni and Re magnetic sublattices in this geometrically frustrated fcc perovskite structure, is revealed as a function of temperature and magnetic field, through the critical behaviour and thermal evolution of microscopic and macroscopic physical quantities.
\end{abstract}
\maketitle

\section{\label{sec:Intro}Introduction}
In the research on multifunctional materials, oxides with a double perovskite structure are continuously attracting our interest due to their structural malleability and a multitude of complex physical properties that arise through their magnetically frustrated geometry \cite{woodward1997octahedral,pardo2009compensated}. Properties such as a magnetoelectric effect, magnetocaloric effect, magnetoresistance or superconductivity may appear in the various phases of perovskite oxides \cite{marx1992metastable,kobayashi1998room,stojanovic2005ferroelectric}. To be able to implement this class of material in future magnetoelectronic and/or spintronic applications, it is of high importance to reveal and understand the static and dynamic processes driven from spin reorientation.\par
\begin{figure*}[htp]
\begin{center}
\hspace{-6mm}
\subfloat[]{\label{fig:1a}\includegraphics[width=75mm,height=53mm]{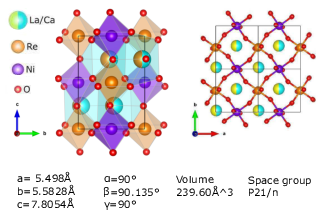}}\hspace{8mm}
\subfloat[]{\hspace{-8mm}\label{fig:1b}\includegraphics[width=75.5mm,height=58.5mm]{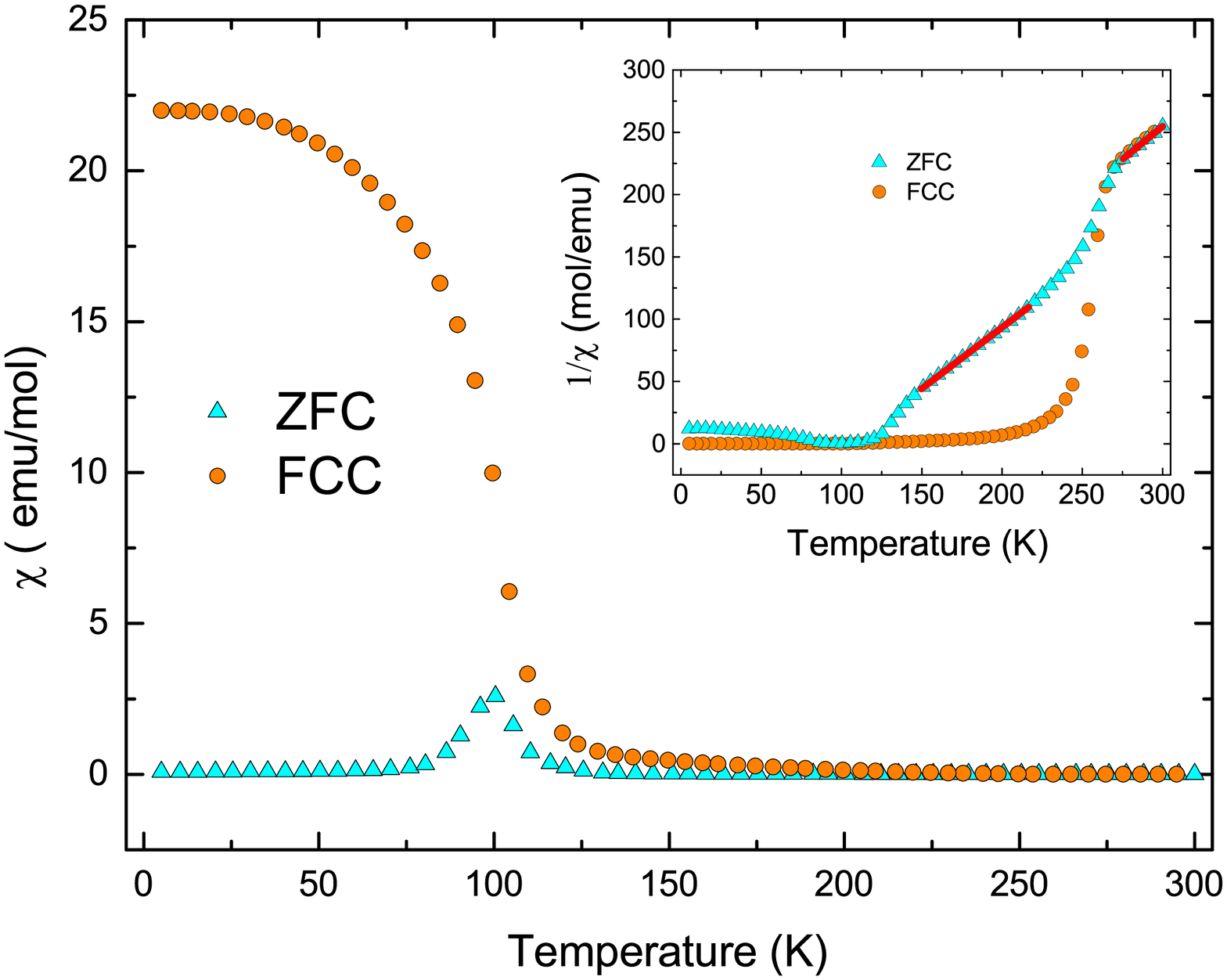}}
\vspace{1mm}\par
\hspace{-4mm}
\subfloat[]{\hspace{-8mm}\label{fig:1c}\includegraphics[width=77.5mm,height=59.1mm]{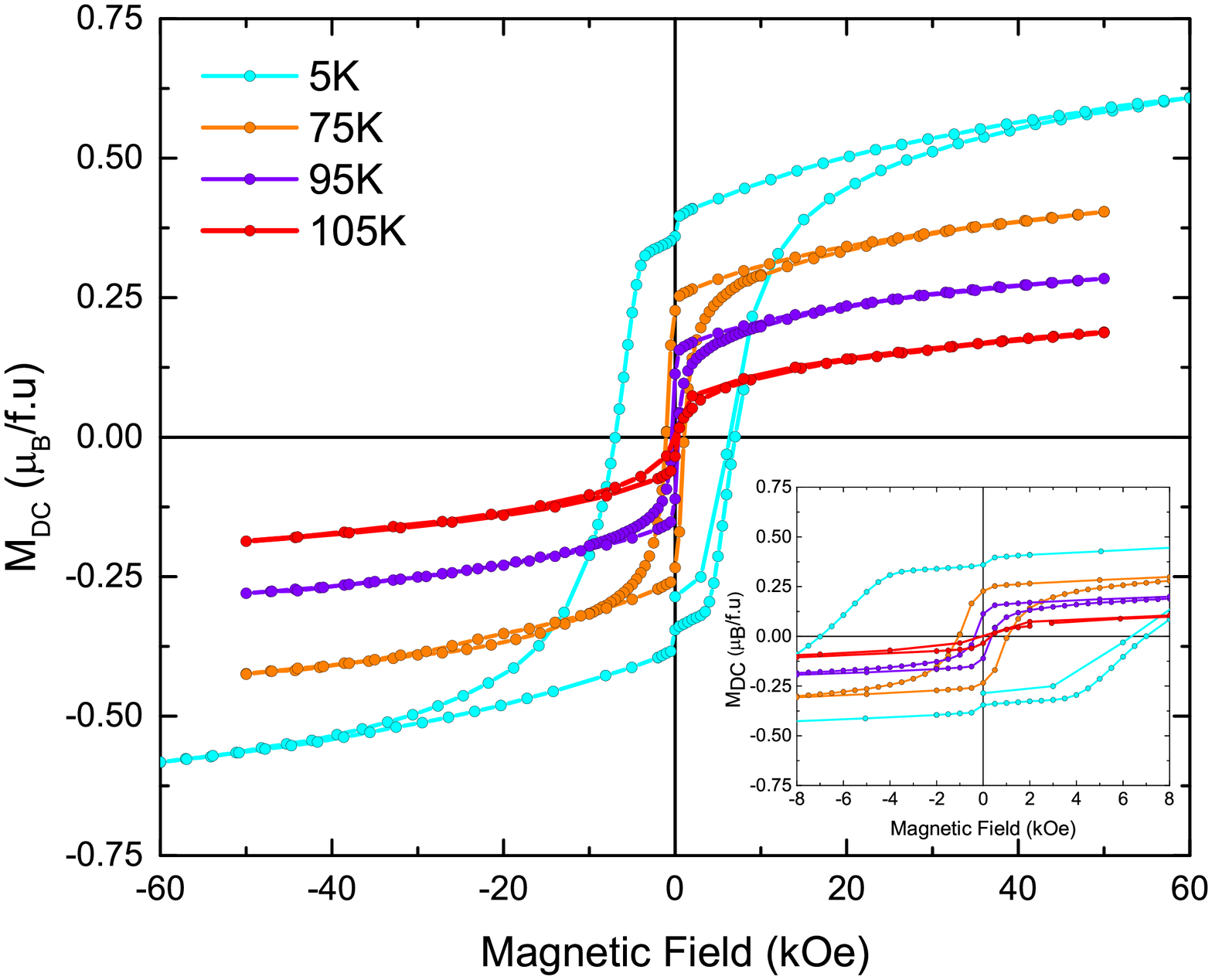}}\hspace{10mm}
\subfloat[]{\hspace{-8mm}\label{fig:1d}\includegraphics[width=75.5mm,height=59.5mm]{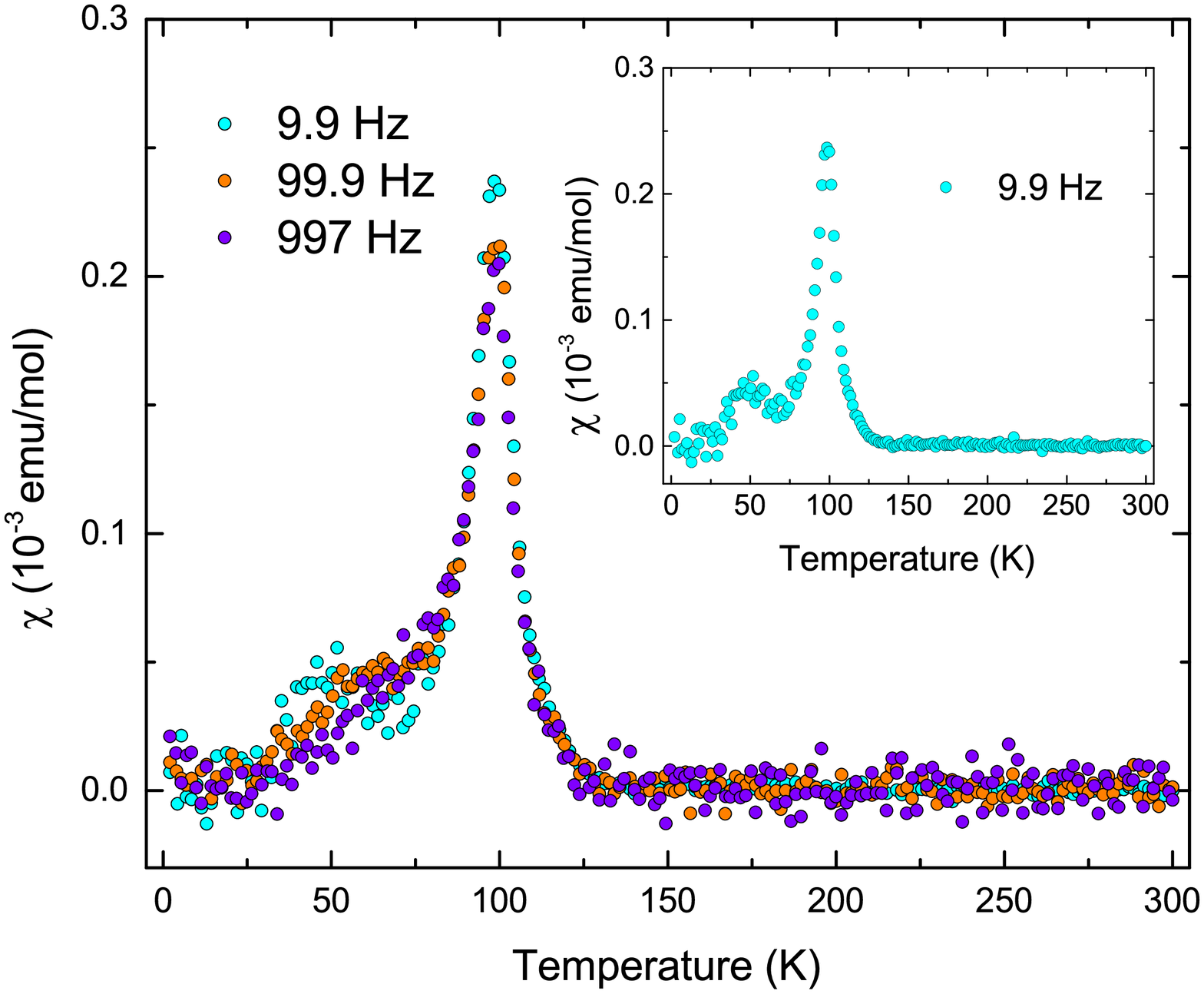}}
  \end{center}
  \vspace{1mm}
  \caption{(a) The crystal structure of LaCaNiReO$_6$, a geometrically frustrated, face-centered cubic, monoclinic crystal lattice that accomodates alternating Ni-O and Re-O octahedra. (b) Temperature dependence of DC magnetic susceptibility in ZFC and FCC sequences. The inset presents the inverse magnetic susceptibility. (c) Magnetic hysteresis loops for $5, 75, 95, 105$~K. The inset focuses around the remanence and coercive field regions. (d) AC susceptibility versus temperature for different frequencies. The evolution of susceptibility at 9.9~Hz is also shown independently for clarity.}
  \label{fig:Fig1}
\end{figure*}

 
The general composition of of double perovskite compounds, illustrated in Fig.~\ref{fig:1a}, is $AA^\prime BB^\prime O_{6}$ where $A,A^\prime$ are alkaline earth or lanthanide cations, and $B,B^\prime$  are 3$d$, 4$d$ or 5$d$ transition metals (TMs) in various oxidation states. The $AA^\prime$ and $B,B^\prime$ 1:1 ratios provide an ordering of $BO_{6},B^\prime O_{6}$ edge-sharing octahedra which form two crystallographically distinct sublattices \cite{singh2010multiferroic,Klein2021}. The structure is flexible to expand/contract, distort or tilt the octahedra due to the Jahn-Teller effect. These distortions are responsible for changes in character of superexchange interactions (ferromagnetic-antiferromagnetic) \cite{jana2019revisiting,aczel2013frustration} or the appearance of Dzyaloshinski-Moriya (DM) interactions \cite{zhu2015strong}, which significantly alter the physical properties of the perovskite \cite{thompson2014long}. The magnetic phases of these compounds are controlled both by the magnetic and non-magnetic cations in the crystal structure. More specifically, the sublattice symmetries, the cation's nominal spin and its spin-orbit coupling (SOC) as well as exchange interactions, are degrees of freedom that determine the magnetic ground state \cite{ding2019magnetic}.\par

A particular interesting double perovskite compound is the LaCaNiReO$_{6}$. This material is the sister compound of LaSrNiReO$_{6}$ \cite{forslund2020intertwined}, and contrary to other double perovskite compound \cite{Shang2018}, the AA’ positions in the crystal structure are occupied randomly by either La$^{3+}$ or Ca$^{2+}$. The BB' positions are also occupied by two different cations, Ni$^{2+}$ and Re$^{5+}$ that, at variance with the A-cations, form an ordered, rock salt-type network or corner-sharing octahedra [Fig.~\ref{fig:1a}]. In double perovskite compounds, the nearest neighbour TM cations are connected by oxygen sites and the superexchange interaction and controls the sign and magnitude of the different magnetic couplings, which depend on both, the TM orbital filling and the B-O-B’ bond angle. According to Goodenough-Kanamori (GKA) rules, the interaction is the strongest at $180^{\circ}$ and is antiferromagnetic if the virtual electron ($e^-$) transfer is between overlapping orbitals that are each half filled, but ferromagnetic if the virtual $e^-$ transfer is from a half filled to an empty orbital or from filled to half filled orbital. At $90^{\circ}$ the interaction is the weakest and it is favoured to be ferromagnetic \cite{kanamori1959superexchange}. For LaCaNiReO$_{6}$, the B-O-B' bonds are bent to an angle of $~152^{\circ}$ at $1$~K \cite{jana2019revisiting}. Between the half filled $e_{g}$ orbitals of Ni$^{2+}$ and partially filled $t_{2g}$ orbitals of Re$^{5+}$, the interaction should favour an antiferromagnetic coupling between nearest neighbours (NN) Ni-Re. In addition there exists a competition between these NN interaction and the longer range, next nearest neighbours (NNN) Ni-Ni and Re-Re. For the $~180^{\circ}$ NNN bonds, according to GKA, a ferromagnetic coupling is possible \cite{ou2018magnetic}. Neutron powder diffraction (NPD) refinement reflects a ground state that is determined by two interacting magnetic sublattices where the Ni spins orient antiparallel to the Re spins, while the moments of each individual sublattice order along the same direction \cite{jana2019revisiting}.\par
In this study we employed both muon spin rotation/relaxation/resonance ($\mu^+$SR) and magnetometry techniques to scrutinize the phases of magnetic ordering in this compound. Muons as local probes can extract space and time dependent information of the material's intrinsic magnetic environment in zero applied fields. A muon spin precession frequency is a direct indication of an ordered state, while the signal relaxation provides information of the ordering range and fluctuations \cite{AlainYaouanc2010}. Since the time window of muons is typically $10^{-12}$ - $10^{-6}$~s , these results are correlated to AC and DC magnetization measurements that provide a bulk view of the sample in a $10^{-6}$ - $1$~s time window, taking into account fast and slow magnetic fluctuations that may occur.\par
Our results show a ferrimagnetic transition to a long-range, commensurate and dynamically ordered state below $T_{\rm N}=103$~K. The ferrimagnetic ordering has been proposed in previous studies \cite{jana2019revisiting}, however, our results suggest that this transition results to a metastable phase. The system is found to undergo a phase transition from a slow to a high dynamics phase between $30-70$~K. The collective relaxation of the spin lattice towards a thermodynamically stable state is characterized by an intrinsic time rate. This defines the magnetization of the system due to an externally applied magnetic field with specific period and amplitude. Above $T_{\rm N}$, the ($\mu^+$SR) fits and magnetization measurements support the existence of two co-existing magnetic phases from $103$~K up to $230$~K, while from $270$~K up to room temperature the paramagnetic phase takes over.

\section{\label{sec:exp} Experimental Details}
A polycrystalline sample of LaCaNiReO$_{6}$ was prepared in stoichiometric ratio by solid state synthesis. La$_2$O$_3$, SrCO$_3$, CaCO$_3$, NiO, Re$_2$O$_7$ and Re metal were used as the starting materials, forming the final compound through mixing, grinding, pelleting and sintering cycles. X-ray and neutron diffraction patterns were recorded at $300$, $125$ and $1$~K to verify the composition. The $P2_1/n$ structural model fits the NPD data at all temperatures, manifesting the absence of a structural phase transition going across the magnetic ordering, based on the resolution of the given NPD data. More information about the synthesis and structure characterization can be found in \cite{jana2019revisiting}.\par
AC-DC magnetization measurements over a $5$-$300$~K temperature range were carried out in a zero-field cooled (ZFC), field-cooled while cooling (FCC) and field-cooled while warming (FCW) process. Magnetic field scans were also performed in a $-60~{\rm kOe}-+60~$~kOe magnetic field range, at various temperatures. The instruments used were a Quantum Design Physical Property Measurement System (PPMS) with a vibration sample magnetometry setup (VSM) as well as  a Magnetic Property Measurement Systems (MPMS) Superconducting Quantum Intereference Device (SQUID).\par
The $\mu^{+}$SR experiments were performed at the surface muon beamline {\bf M20} in the TRIUMF facilities. A 100\% polarized, continuous, positive muon beam was targeted onto an aluminium coated mylar envelope of approximately 1~cm$^2$ surface, filled with $\sim$1~g of the powdered sample. The sample was inserted in a $^{4}$He cryostat and reached a $2$~K base temperature. The software package $musrfit$ was used to analyze the data \cite{suter2012musrfit}. Measurements were performed in the zero field (ZF), longitudinal field (LF) and transverse field (TF) geometry, with respect to the initial muon spin polarization. Muons are implanted one at a time into the sample, and come to rest at an interstitial site. There, muons interact with the local magnetic field, undergoing a Larmor precession and finally emitting a positron, with a high probability in the direction of the spin orientation before decay \cite{AlainYaouanc2010}. Our data sets consist of $\sim$10 million positron counts for the TF and LF measurements, and $\sim$35 million counts for the ZF measurements.

\begin{figure}[htp]
  \begin{center}
  	\subfloat[]{\hspace{-8mm}\label{fig:2a}\includegraphics[width=71.5mm,height=55mm]{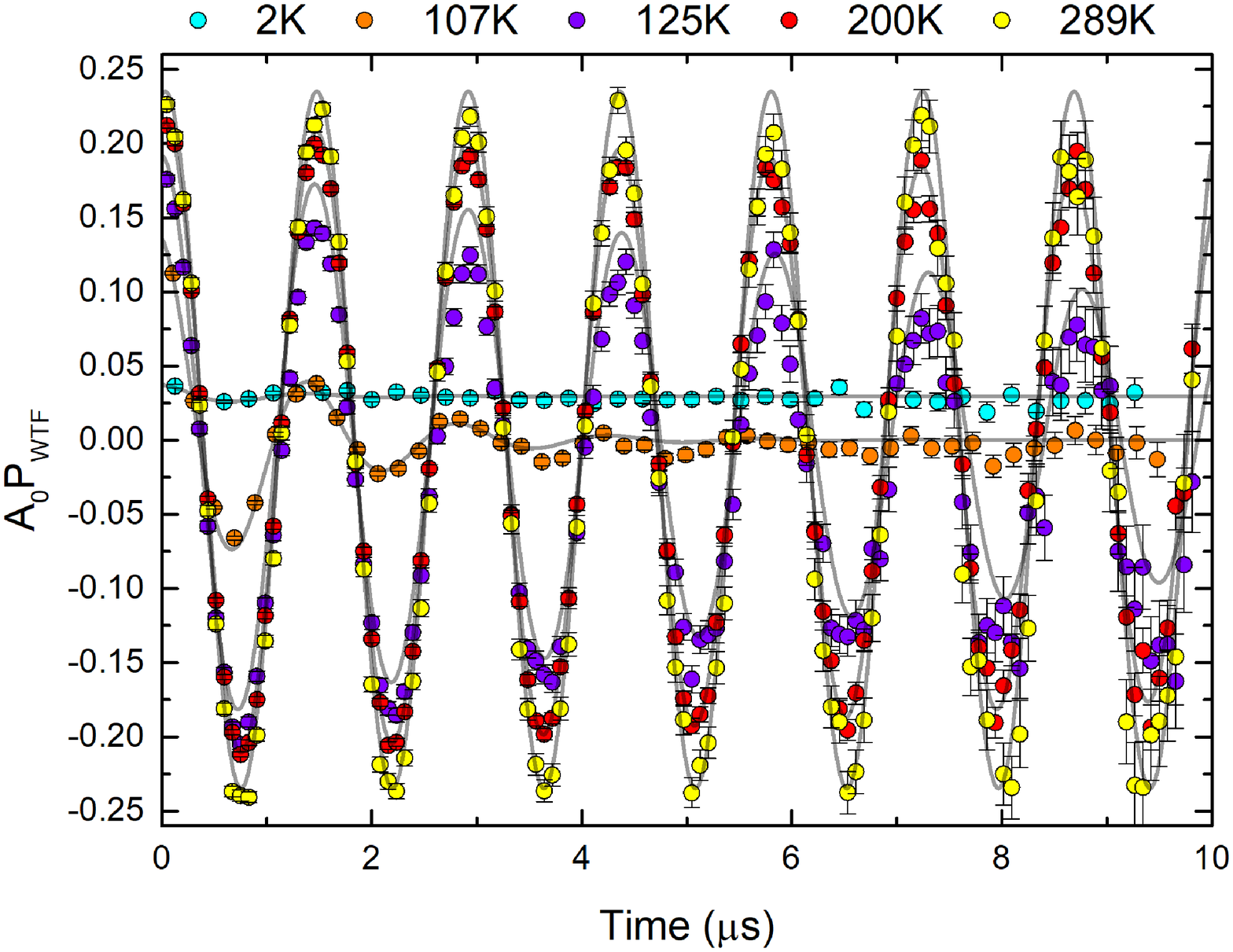}}\quad
  	\vspace{1mm}
  	\subfloat[]{\hspace{-8mm}\label{fig:2b}\includegraphics[width=72.5mm,height=55mm]{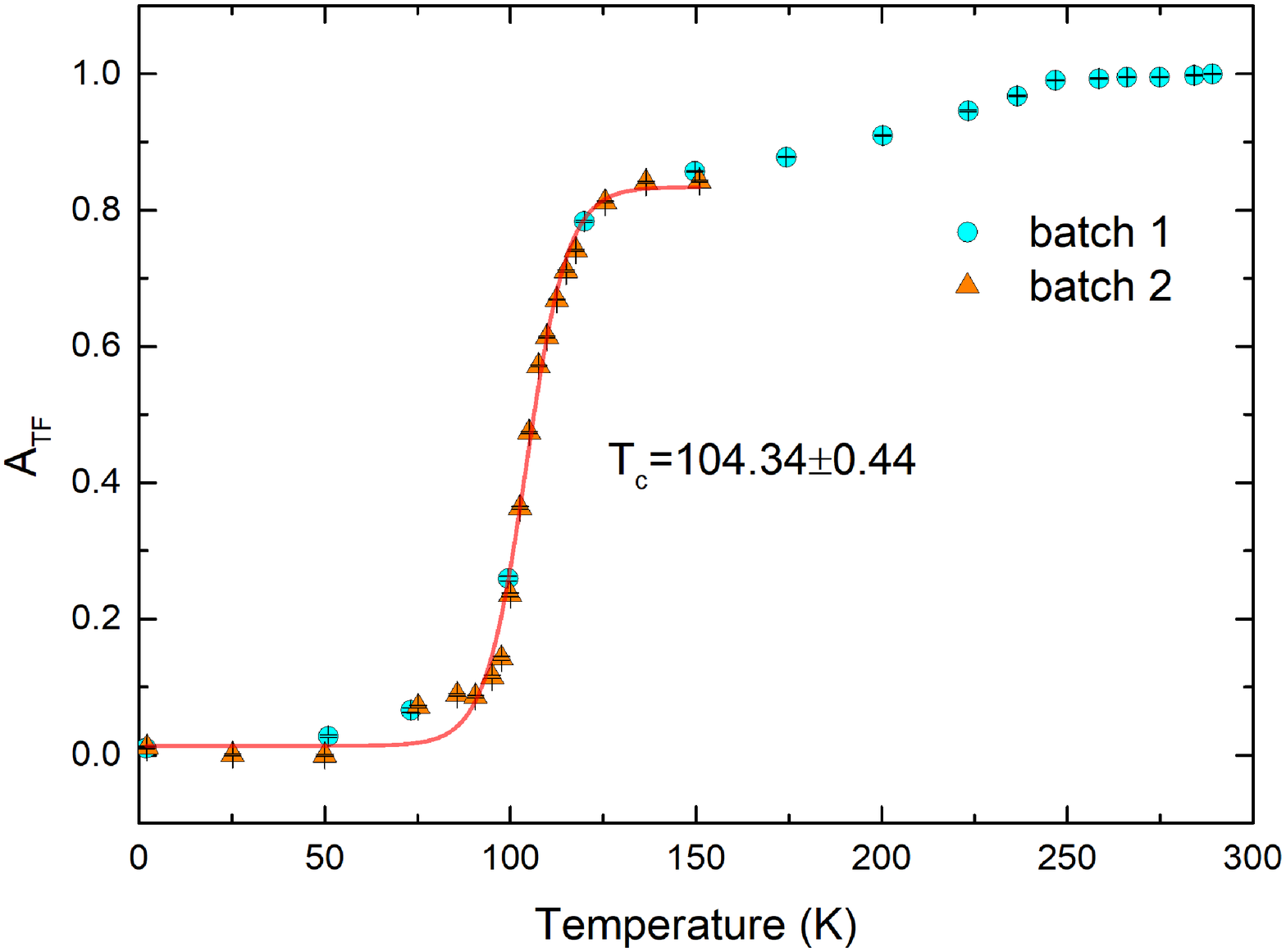}}\quad
  	\vspace{1mm}
	\subfloat[]{\hspace{-8mm}\label{fig:2c}\includegraphics[width=73.5mm,height=55mm]{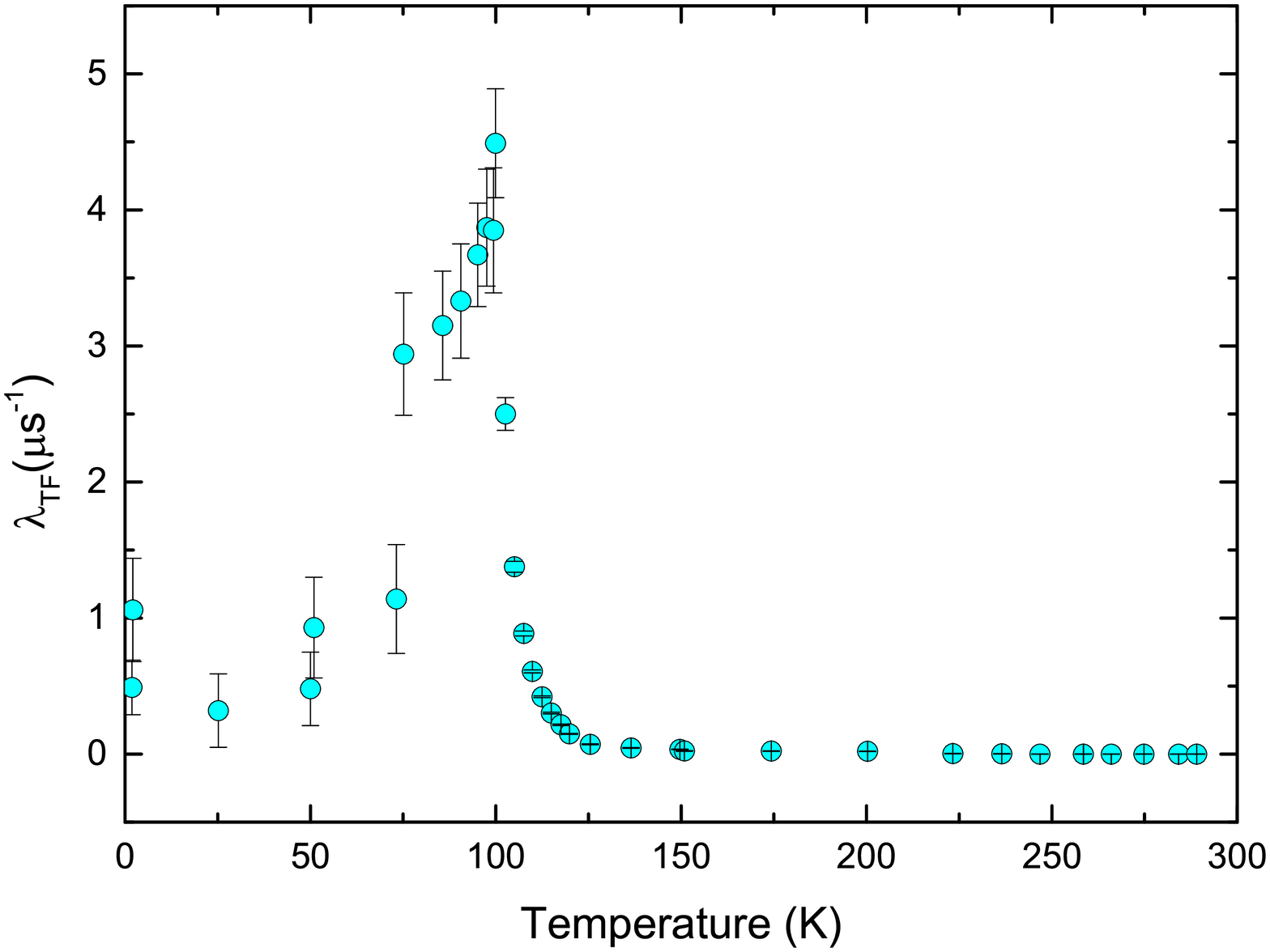}}\quad
    \end{center}
    \vspace{1mm}
  \caption{(a) Weak transverse field (TF$=50$~Oe) time spectra recorded at $T$= 45, 107, 125, 200 and 289~K. The solid lines are fits obtained from Eq.(\ref{eq:wTF}).The oscillation amplitude clearly increases with increasing temperature, although the maximum asymmetry is fully recovered at $289$~K. (b) The asymmentry [Eq.(\ref{eq:wTF})] versus temperature shows a magnetic transition at $104.3(4)$~K taking place in two stages, between $130$-$90$K and $90$-$50$~K. A second transition appears also above $230$~K. (c) The relaxation of the oscillatory component exhibits a peak around the transition temperature while a shoulder appears at $75$~K.
  }
  \label{fig:Fig2}
\end{figure}

\section{\label{sec:results}Results}
\subsection{Magnetic Susceptibility}
Temperature dependent DC magnetic susceptibility was measured at $5$-$300$~K with an applied field of $100$~Oe, as shown in Fig.\ref{fig:1b}. The FCC and FCW (not displayed) curves were identical meaning that there was no quenching of magnetic moments during the cooling. The bifurcation between the FCC and ZFC curves reveals a change in the magnetic response of the system around $213$~K. The FCC susceptibility at low temperatures reveals a ferromagnetic component and from its first derivative we determined a Curie temperature $T_{\rm C}=100.3(3)$~K. To extend the results of the previous study \cite{jana2019revisiting} below the Curie temperature, we show explicitly that the ZFC susceptibility reaches the value of $0.08~$~emu/mol at $5$K for an 100Oe applied field. The ZFC susceptibility increases with increasing temperature and obtains a maximum close to the derived Curie temperature. The ZFC and FCC inverse susceptibility is presented in Fig.\ref{fig:1b}. The region above transition could not be fitted with a hyberbola according to Neel's molecular field model \cite{smart1955neel}. Consequently, we fitted the two linear regions (highlighted red regions) at $150~{\rm K}<T<220~{\rm K}$ and $270~{\rm K}<T<300~{\rm K}$ to the Curie-Weiss law $\frac{1}{\chi}=\frac{T-\theta}{C}$ \cite{de2010valence}. The high temperature fit resulted to a Curie constant of $C_h=0.94(2)$~K and Curie temperature $\theta_h=60(4)$~K, while the low temperature fit to $C_l=1.06(3)$~K and $\theta_l=100(2)$~K. The ZFC-FCC curve types below transition and the calculated Curie temperature values are characteristic features of a ferrimagnetic ordering. The effective magnetic moments were then calculated from $\mu_{\rm eff}=2.83(\frac{C_m}{Z})^\frac{1}{2} \mu_{\rm B}$, where $C_m$ the molar Curie constant, $Z$ the formula unit per unit cell and $\mu_B$ the Bohr magneton \cite{yang2007structural}. The respective effective moments are $\mu_{\rm eff/h}=0.63~\mu_{\rm B}$ and $\mu_{\rm eff/l}=0.67~\mu_{\rm B}$. If we consider unquenched or partially quenched orbital moments for the Re and Ni ions in the octahedral complexes, the total $\mu_{\rm eff}$ is given by $\mu^2_{\rm eff}=xg^2J(J+1)\mu^2_{\rm B}$, where $x$ the fraction of magnetic ions per formula unit, $g$ their gyromagnetic factor and $J$ the total angular momentum. Assuming that the Ni$^{2+}$ and Re$^{5+}$ magnetic systems have the same ordering temperature through superexchange coupling \cite{de1999magnetic,gupta2019site}, we calculate $\mu_{\rm eff}=\sqrt{\mu^2_{\rm eff.Ni}-\mu^2_{\rm eff.Re}}=1.03~\mu_{\rm B}$, where $\mu_{\rm eff.Ni}=1.81~\mu_{\rm B}$ and $\mu_{\rm eff.Re}=1.49~\mu_{\rm B}$. This theoretical value is an overestimation to the experimentally extracted values from our magnetization measurements, but still in good agreement with the values obtained from powder neutron diffraction experiments \cite{jana2019revisiting}. The precision of the spin determination of the magnetic moment is limited since SOC in Re ions, Jahn-Teller distortions of Re-O octahedra, and the possible canted magnetic structure affect the spontaneous magnetization of the compound.\par
Isothermal field dependent DC magnetization measurements were performed at $5$, $75$, $95$, $105$ and $300$~K for magnetic fields from -$60$~kOe to +$60$~kOe. The evolution of the magnetization is presented in figure Fig.\ref{fig:1c} (the paramagnetic, linear behaviour at $300$~K is not shown). Magnetisation saturation is never reached for ferrimagnets, however at the maximum field of $60$kOe we consider an effective magnetic moment of $\mu_{\rm eff5K}=0.60~\mu_{\rm B}$ at $5$~K, following the previous experimental values. A long range order is evident and a very large coercive field ($\approx7$~kOe) is measured. High coercivity and remanence has been also observed in other Re-based double perovskites \cite{Teresa2004, Kato2002, Alamelu2002} and is attributed to an intrinsic anisotropy of these compounds. Taking into account a strong SOC of Re ions, first-principle calculations \cite{Jeng2003} predict a large unquenched orbital moment, which is thought to be the origin of the magnetic anisotropy \cite{Bloch1931}. In the $5$~K hysteresis loop, shoulders appear at zero field which indicate spin reorientation \cite{ahmed2015magnetic,heisz1987effect}. In the recorded loops at $75$~K and above, these shoulders disappear, while the remanence and coercivity approach zero at $105$~K.\par
AC magnetic susceptibility measurements were also performed at three ac field frequencies, to probe the magnetic relaxation in the two sublattices spin system as shown in Fig.~\ref{fig:1d}. The ordering temperature does not depend on frequency, which excludes a spin glass behaviour. We observe for the first time that a second peak in the susceptibility curve rises at $50$~K as the frequency is tuned from 1000~Hz to 10~Hz [see inset Fig. \ref{fig:1d}]. This can be an effect of reordering of spin domains with a subtle equilibrium, that respond to an ac field. This fluctuation is another indication of reorientation and freezing of transverse spin components \cite{saito1986mictomagnetism}.

\subsection{Muon Spin Rotation}
The $\mu^+$SR study consists of weak TF, ZF and LF measurements. We can extract the local field distribution from the time spectra of the muon decay asymmetries between the surrounding detectors \cite{de1997muon,garwin1957observations}. From the TF spectra we extract the initial and baseline asymmetry which are used as constants when treating the ZF time spectra. The ZF and LF measurements follow in order to gain detailed information about the conditions of magnetic ordering, as well as the intrinsic magnetic fields, arising from nuclear and electronic moments.

\begin{figure}[htp]
  \begin{center}
  	\subfloat[]{\hspace{-6mm}\label{fig:3a}\includegraphics[width=75.5mm,height=55mm]{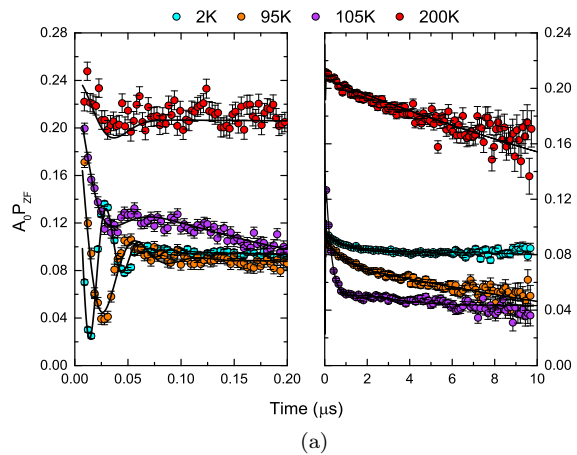}}\quad
  	\vspace{1mm}
  	\subfloat[]{\hspace{-8mm}\label{fig:3b}\includegraphics[width=73.5mm,height=55mm]{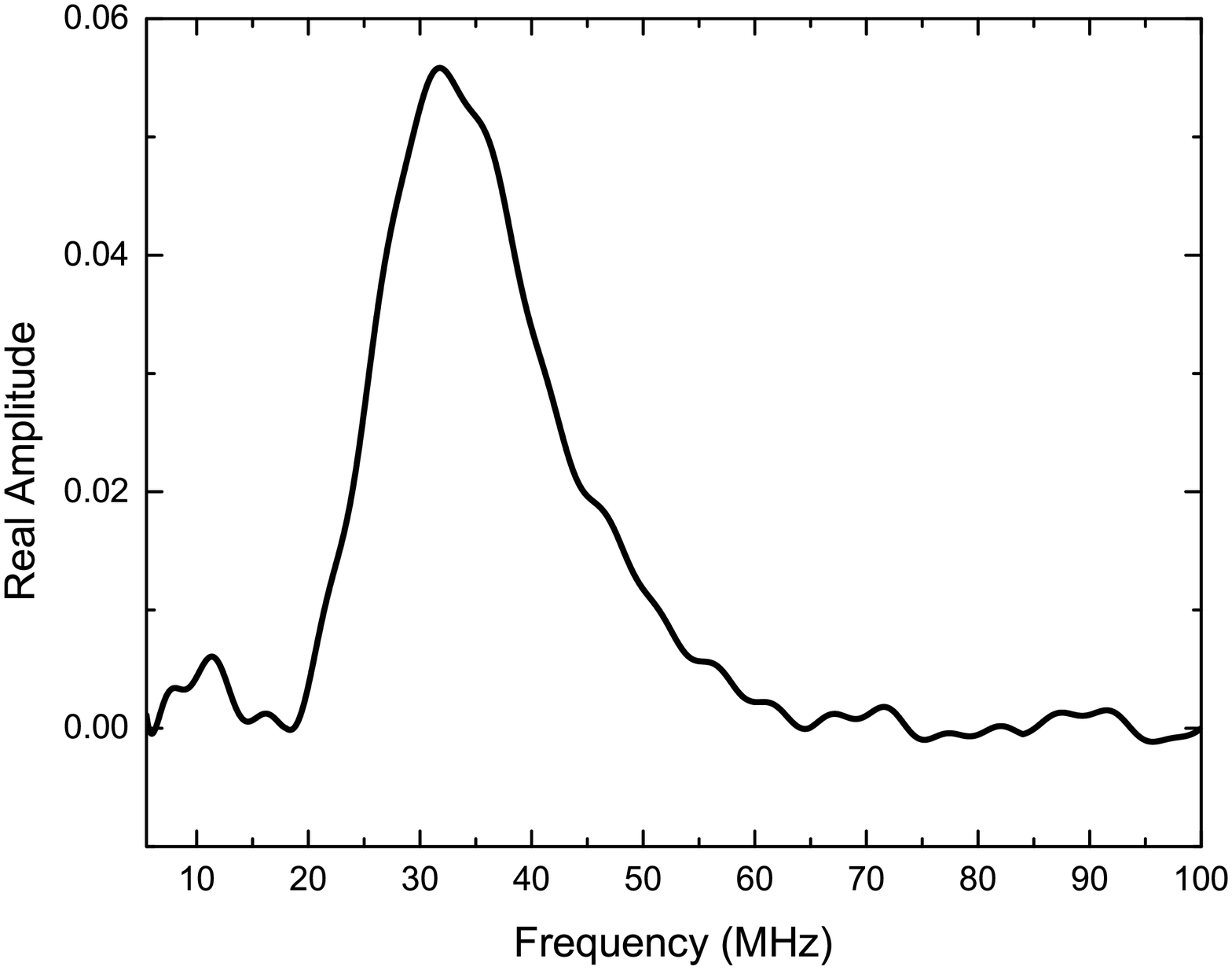}}\quad
  	\vspace{1mm}
	\subfloat[]{\hspace{-8mm}\label{fig:3c}\includegraphics[width=70mm,height=55mm]{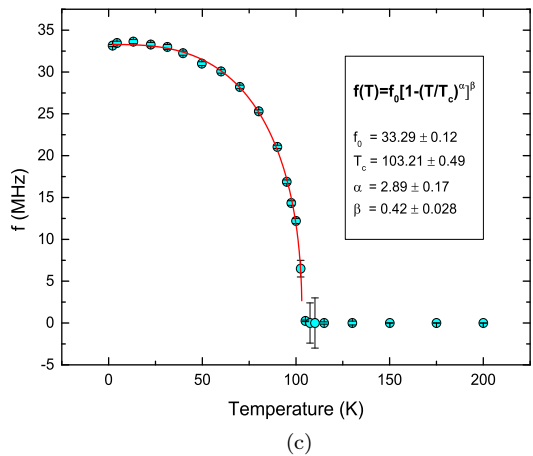}}\quad
  \end{center}
  \vspace{1mm}
  \caption{(a) Zero field time spectra at $2$, $95$, $105$ and $200$~K fitted with Eq.~(\ref{eq:ZF}). The left and right window present the short and long time domain of the ZF time spectrum, respectively. A damped oscillation is observed up to $0.20~\mu$s  while the longer time domain displays a exponential relaxation. (b) The real part of the Fourier transform which exhibits the oscillation frequency at 2~K. (c) The frequency component as a function of temperature, fitted with the power law function \ref{eq:magn}. A magnetic transition appears at $T_{\rm N}=103.2(5)$~K where the oscillation of the signal takes effect.}
  \label{fig:Fig3}
\end{figure}
\subsubsection{Weak transverse field (TF)}
Muon depolarization spectra were recorded for ${\rm TF}=50$~Oe in the temperature range $2~{\rm K}\leq T \leq300~{\rm K}$. We present the evolution of the TF spectra at selected temperatures, above and below the magnetic phase transition in Fig.~\ref{fig:2a}. Following the oscillation that is driven from the applied external magnetic field, the TF time spectra were fitted using an oscillatory component multiplied by a non-oscillatory depolarizing component according to:

\begin{eqnarray}
 A_0 \, P_{\rm TF}(t) &=&
A_{\rm TF} \cos(2\pi ft+\phi)e^{-(\lambda t)}+A_{\rm bg}
\label{eq:wTF}
\end{eqnarray}

where $A_{0}$ is the initial asymmetry and $P_{\rm TF}$ is the muon spin polarisation function. $A_{\rm TF}$, $f$, $\phi$ and $\lambda$ are the asymmetry, frequency, relative phase and depolarisation rate of the implanted muons under applied TF, while $A_{\rm bg}$ accounts for background muon depolarization. We observe a dumping of the oscillation with decreasing temperature, towards a magnetically ordered state, with an internal field that also takes part, together with the applied field, in the muon spin precession. The oscillation frequency is proportional to the applied magnetic field while the asymmetry $A_{\rm TF}$ corresponds to the externally magnetized fraction of the sample. The transverse field asymmetry is displayed as a function of temperature in figure Fig.~\ref{fig:2b} and is fitted with a Boltzmann sigmoid function which outputs a transition temperature $T_{\rm c}=104.3(4)$~K. A step appears below transition, between $100$-$50$~K, indicating that the static magnetic ground state is not yet reached.
Above transition, the sample enters the paramagnetic phase and full asymmetry is recovered only above $230$~K, a behaviour that corroborates the dc susceptibility results. The depolarization rate $\lambda_{\rm TF}$ is presented in figure Fig.~\ref{fig:2c}. At high temperatures above transition, the damping is approaching zero as the fluctuation of local magnetic moments increases resulting to a motional narrowing effect. When temperature is decreased, $\lambda_{\rm TF}$ peaks around the magnetic phase transition, at the critical slowing down of magnetic fluctuations. At $75$~K a shoulder appears, pointing to the dynamic phase transition also observed in AC susceptibility. Eventually, a static, commensurate magnetic ordering appears to be achieved at base temperature.

\subsubsection{Zero field (ZF)}
ZF measurements were carried out in the temperature range $2~{\rm K}\leq T \leq200~{\rm K}$. We present a selection of muon spin depolarization spectra at temperatures above and below the magnetic transition in Fig.~\ref{fig:3a}. In the short time scale, the onset of a precession and a fast relaxation of the muon spin ensemble appears below $100$~K. This damped oscillation is the effect of a static internal field distribution, perpendicular to the muon spin. At base temperature $T=2$~K an oscillation appears with a single precessing frequency at 33.2~MHz up to $0.1~\mu$s, which denotes a commensurate magnetic ordering. In the long time scale, a slower relaxation describes the spin dynamics which correspond to the longitudinal field components to the muon spin.\hfill \break

\paragraph{Below the transition temperature.}
\hfill \break \par
The time spectrum is fitted with an internal field \cite{Blundell_1995} , oscillating function and a Lorentzian Kubo-Toyabe (LKT) function \cite{umar2021muon} up to $105$~K:

\begin{eqnarray}
 A_0 \, P_{\rm ZF}(t) &=&
 A_{\rm IF}\left [ \xi \cos (2\pi ft+\varphi) e^{-\lambda_{\rm T}t}+(1-\xi)e^{-\lambda_{\rm L}t}\right ]\cr &+&
A_{\rm LKT}\left [   \frac{1}{3}+\frac{2}{3}(1-\Delta t)e^{-(\Delta t)}\right ] 
\label{eq:ZF}
\end{eqnarray}

where $A_{\rm IF}$ is the asymmetry, $f$ the frequency which is a measure of the sublattice magnetization, and $\phi$ the relative phase which is set to zero. The $\lambda_{\rm T}$ and $\lambda_{\rm L}$ are the transverse and longitudinal depolarization rate applied to the fast precessing part and slow relaxing tail respectively. In the KT function each orthogonal component of the magnetic field at the muon site is represented by a probability distribution width $\Delta$, the corresponding distribution width.\par

\begin{figure}[!htp]
  \begin{center}
\subfloat[]{\hspace{-8mm}\label{fig:4a}\includegraphics[width=72.5mm,height=55mm]{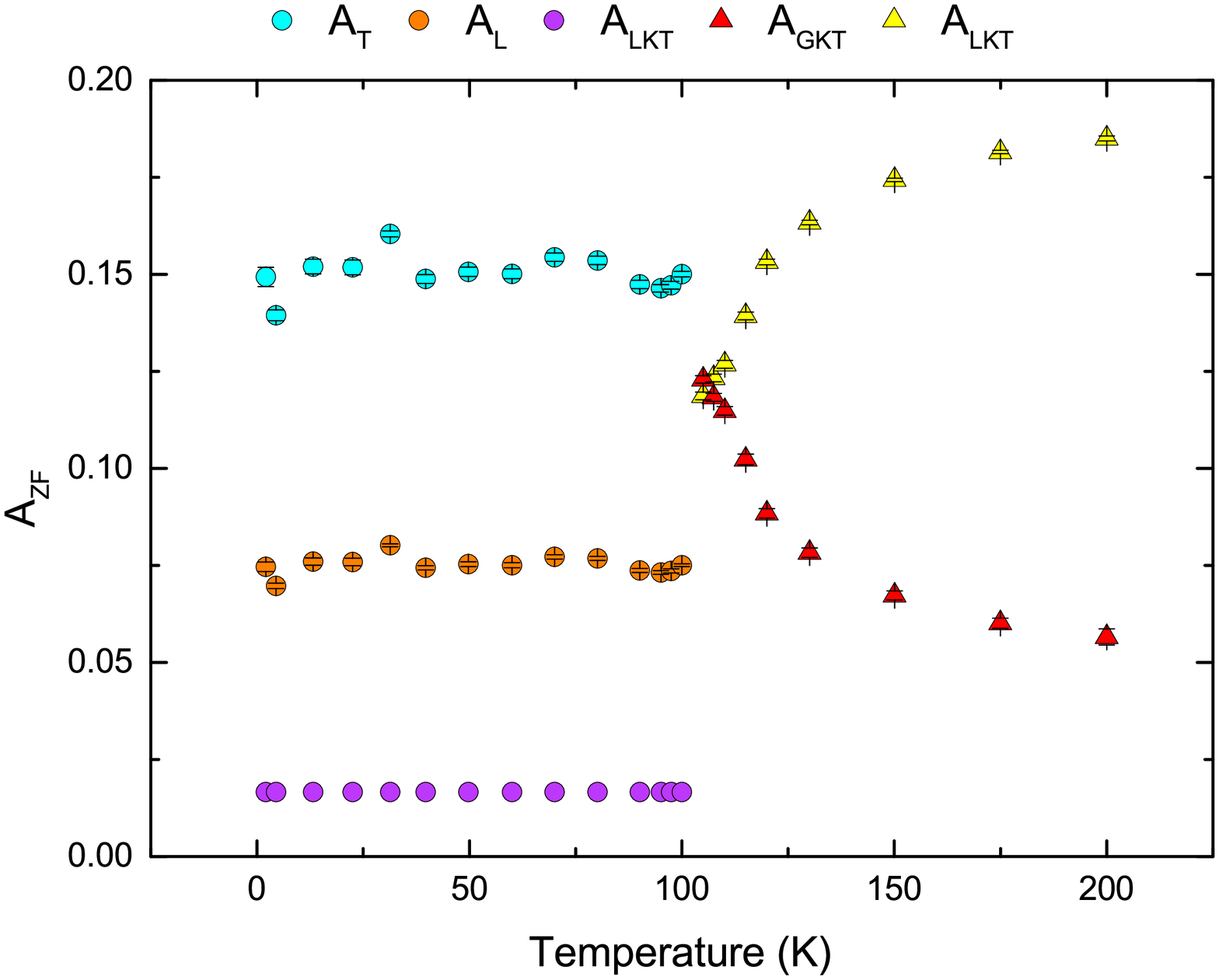}}\quad
\vspace{1mm}
\subfloat[]{\hspace{-8mm}\label{fig:4b}\includegraphics[width=75mm,height=55mm]{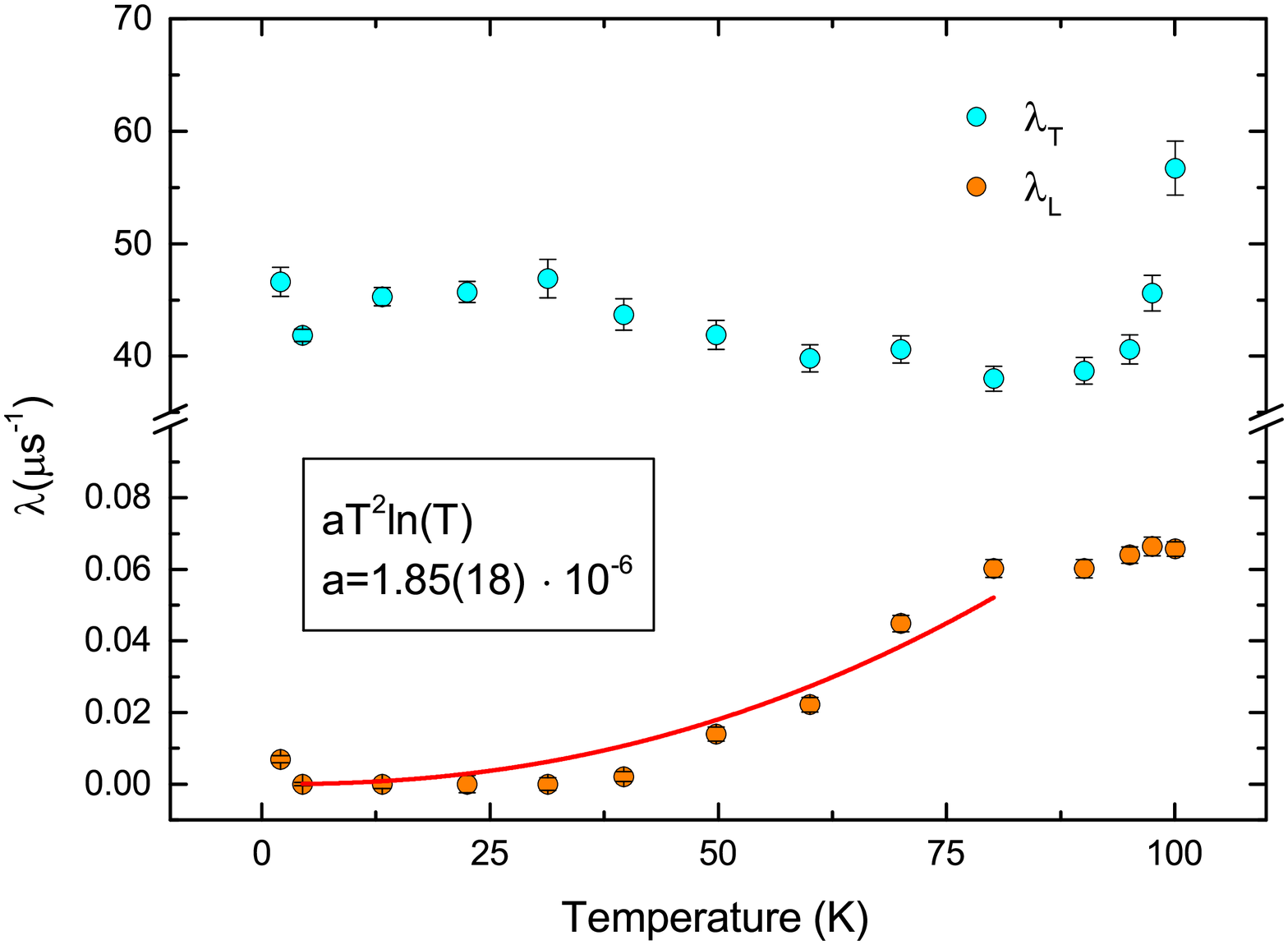}}\quad
\vspace{1mm}
\subfloat[]{\hspace{-8mm}\label{fig:4c}\includegraphics[width=75mm,height=55mm]{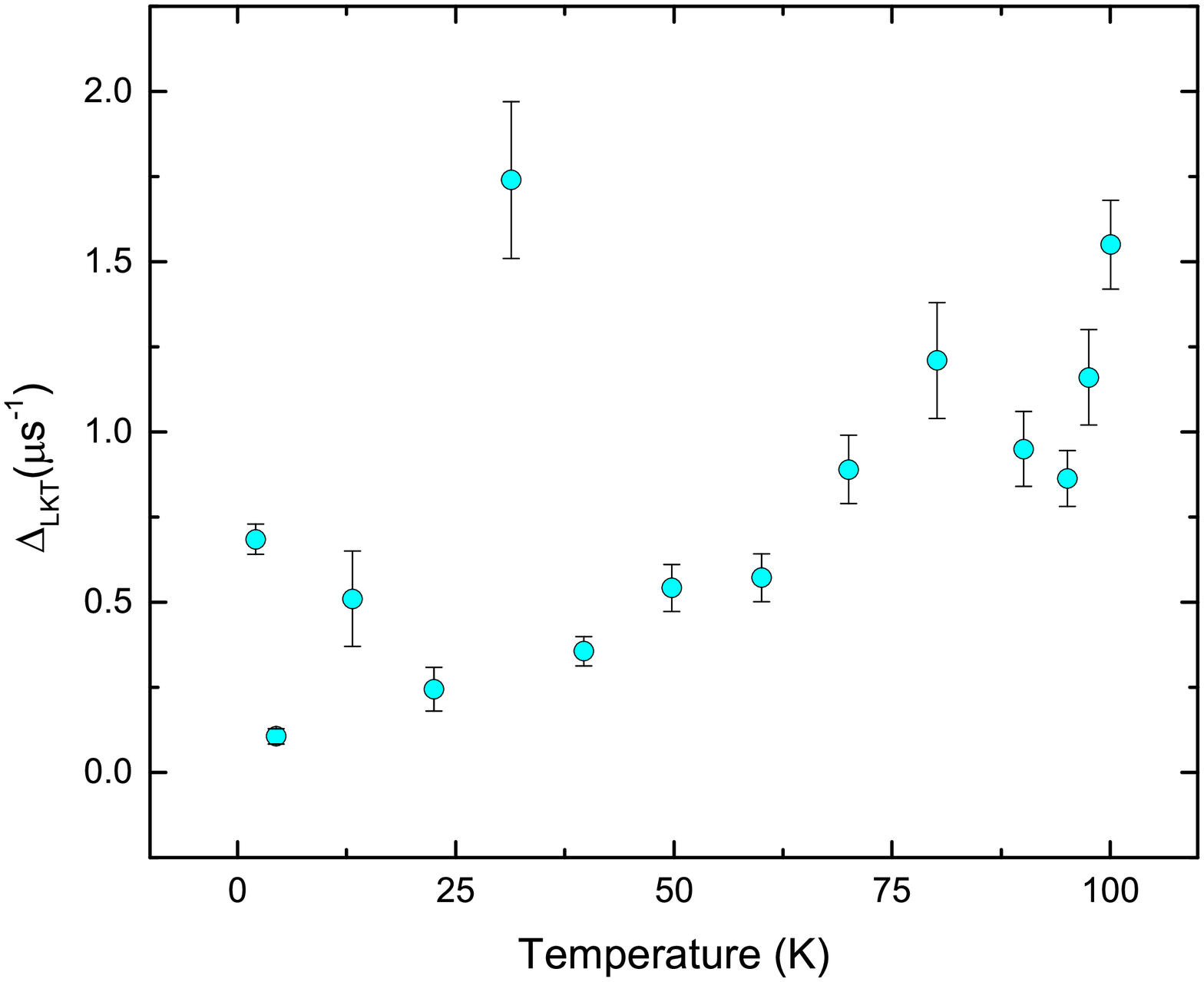}}\quad
  \end{center}
  \vspace{1mm}
  \caption{Fit results for the zero field data fitted with Eq.~(\ref{eq:ZF}). All fit parameters are presented as a function of temperature: (a) Asymmetries for the depolarization contributions to the muon ensembles. (b) The fast $\lambda_{\rm T}$ and slow $\lambda_{\rm L}$ depolarization rates. (c) Field distribution width of the Lorentzian KT.}
  \label{fig:Fig4}
\end{figure}
From DFT calculations on the sister compound LaSrNiReO$_6$ two possible muon sites were predicted \cite{forslund2020intertwined}. For the LaCaNiReO$_6$ compound one frequency deemed appropriate to reconstruct and analyse the data. A Bessel function or the use of two oscillators did not produce a consistent fitting. The use of a single oscillation frequency implies that the muon sites appear to be equivalent below the transition temperature. This precession frequency is indicated in the real part of the Fourier transform of the time spectrum, as shown in Fig~\ref{fig:3b}.  In Fig.~\ref{fig:3c} this frequency, as extracted from the ZF data, is presented against temperature. The frequency is proportional to the sample magnetization and the experimental data can be fitted by a power-law function \cite{thompson2014long,pelka2013magnetic,AlainYaouanc2010} that considers both spontaneous magnetization and spin-wave excitations at low temperatures, but also the magnetic anisotropy that becomes significant near the Curie temperature:
\begin{eqnarray}
f(T)=f_{0}\left [ 1-\left ( \frac{T}{T_{\rm N}} \right )^{\alpha } \right ]^{\beta }
\label{eq:magn}
\end{eqnarray}
where $f_0$ is proportional to the spontaneous magnetization at base temperature. The critical exponent $\alpha$ corresponds to the low temperature properties, and $\beta$ determines the asymptotic behaviour near the transition temperature. The fit results to a critical temperature $T_{\rm N}=103.2(5)$~K. This is the actual critical temperature of the sample since the ZF measurement probes the intrinsic magnetic ordering without excitations. The critical exponent $\beta=0.42(3)$ matches to the mean-field model, while the low temperature exponent $\alpha=2.89(17)$ follows the $T^{5/2}$ power law according to the Dyson formalism \cite{Mattis2012}. This behaviour describes a system of two spin waves interacting in a ferromagnet. In case of a ferrimagnet, the magnon excitation includes transverse fluctuations of both sublattice spins and the dispersion curve consists of two branches \cite{karchev2008towards,kaplan1952spin}.\par 
\begin{figure}[t]
  \begin{center}
\subfloat[]{\hspace{1mm}\label{fig:5a}\includegraphics[width=81mm,height=56mm]{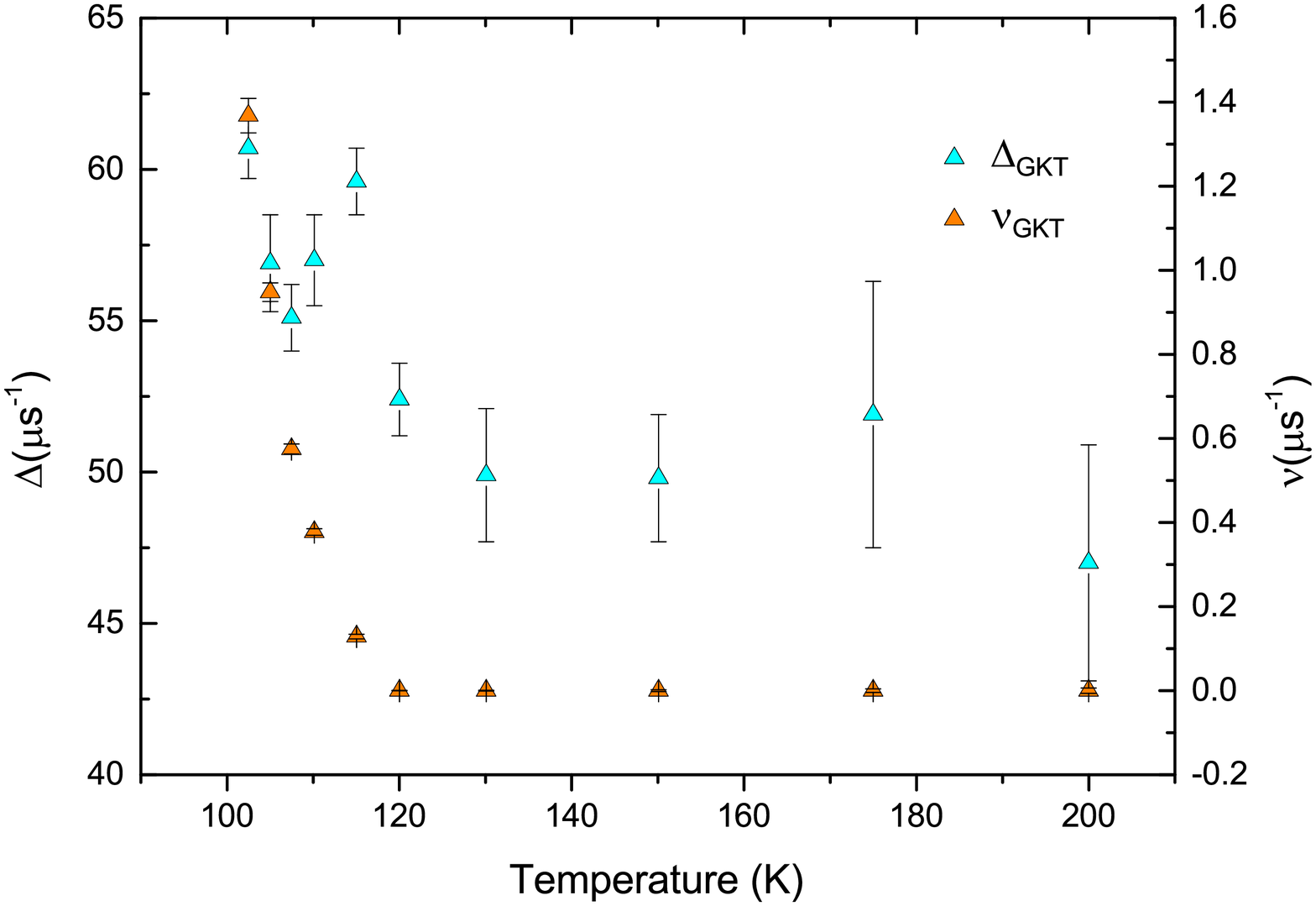}}\quad
\vspace{1mm}
\subfloat[]{\hspace{2mm}\label{fig:5b}\includegraphics[width=81mm,height=56mm]{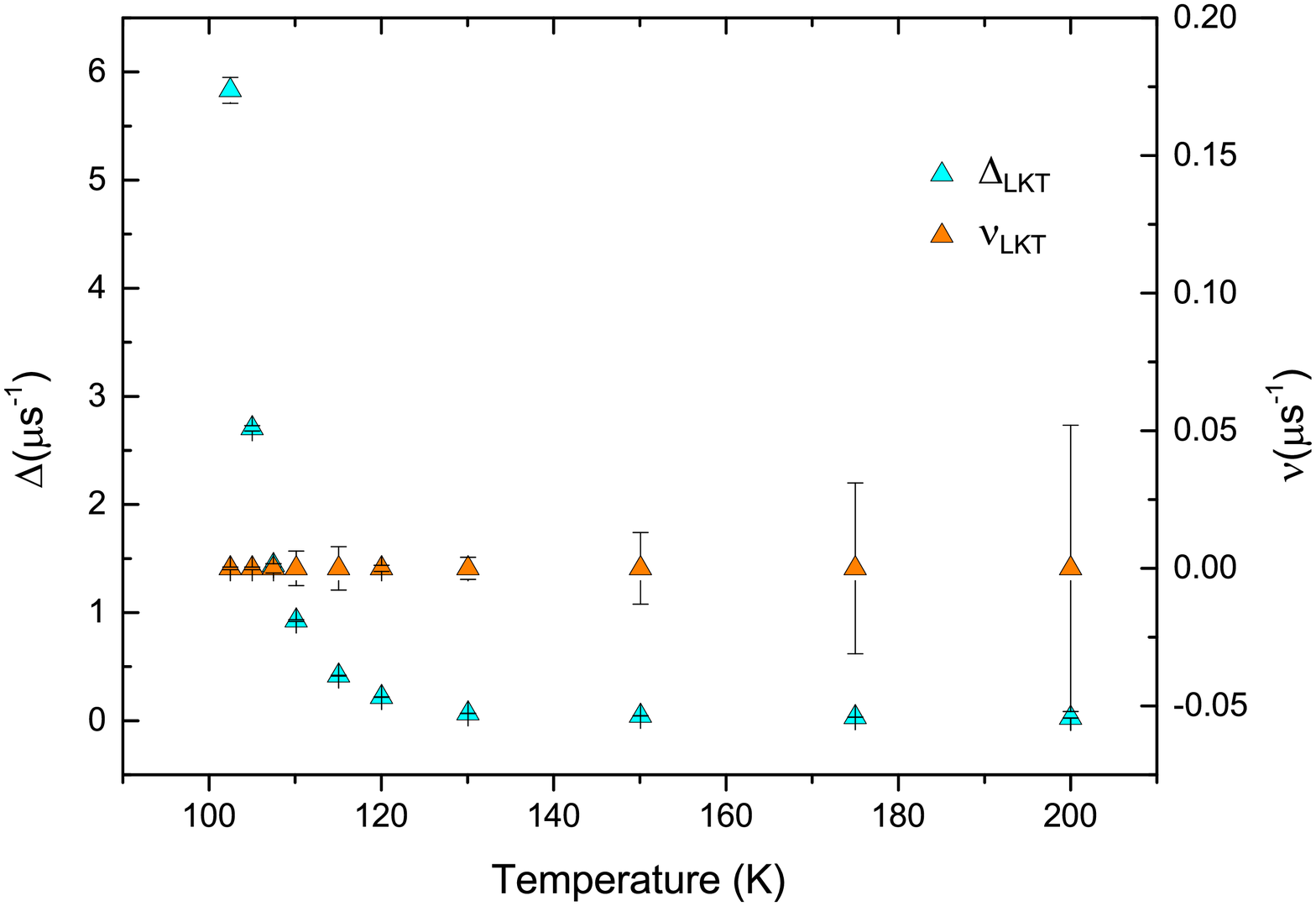}}\quad
  \end{center}
  \vspace{0.5mm}
  \caption{The temperature dependencies of the dynamic Kubo-Toyabe parameters of Eq.~(\ref{eq:ZFh}) in the temperature range $102.5-200$~K. The field distribution width and field fluctuation rate (a) for the Gaussian KT and (b) for the Lorentzian KT.}
  \label{fig:Fig5}
\end{figure}
The parameter $\xi$ in Eq.(\ref{eq:ZF}) is used to regulate the $A_{\rm T}=\frac{2A_{\rm IF}}{3}$ and $A_{\rm L}=\frac{A_{\rm IF}}{3}$ ratio, describing the perpendicular (precessing) and the parallel (relaxing) tail components. Below transition both $A_{\rm T}$, $A_{\rm L}$ and the contribution from the KT function are constant as illustrated in figure Fig.~\ref{fig:4a}. The $A_{\rm LKT}$ corresponds to a $6\%$ volumic fraction of the sample and the field distribution width narrows down with lower temperatures. This contribution is attributed to randomly oriented dilute spins. Both depolarization rates $\lambda_{\rm T}$, $\lambda_{\rm L}$ increase as the temperature approaches the transition, revealing an increase in dynamics. The transverse component has a high value down to base temperature since it depends on the static field distribution [Fig.~\ref{fig:4b}]. The longitudinal component on the other hand slowly approaches zero as the spin dynamics weaken but are still existent [Fig.~\ref{fig:4c}]. Both $\lambda_{\rm T}$ and $\lambda_{\rm L}$ denote a dynamic phase between $30~{\rm K}<T<100~{\rm K}$ in agreement with the TF results and the bulk susceptibility measurements. A reasonable interpretation can be to associate these features with the slowing down of small, canted components of the spin sublattices. In this region the spin-lattice is relaxing until it reaches a static, commensurate phase below $30$~K. The evolution of $\lambda_{\rm L}$ rate, a measure of spin lattice relaxation, can be described as a function of temperature $aT^2ln(T)$ below $T_{\rm N}$. This dependence has been identified to be a signature of muon depolarization from a two magnon excitation \cite{gubbens1994study,beeman1968nuclear}, consistent with the power law magnetization model.\hfill \break
\paragraph{Above the transition temperature.}
\hfill \break\par
From $102.5$~K to $200$~K the ZF-$\mu^+$SR spectra were fitted with a combination of a dynamic Gaussian Kubo-Toyabe function ($G_{\rm KT}^{dyn}$) and a dynamic Lorentzian Kubo-Toyabe function ($L_{\rm KT}^{dyn}$)  \cite{feyerherm1997competition}:
\begin{eqnarray}
A_0\,P_{\rm ZF}(t)=A_{\rm G}G_{\rm KT}^{dyn}(\Delta,\nu,t)+A_{\rm L}L_{\rm KT}^{dyn}(\Delta,\nu,t)
\label{eq:ZFh}
\end{eqnarray}

\begin{figure*}[ht]
\begin{center}
\hspace{-6mm}
\subfloat[]{\hspace{-8mm}\label{fig:6a}\includegraphics[width=75mm,height=57.5mm]{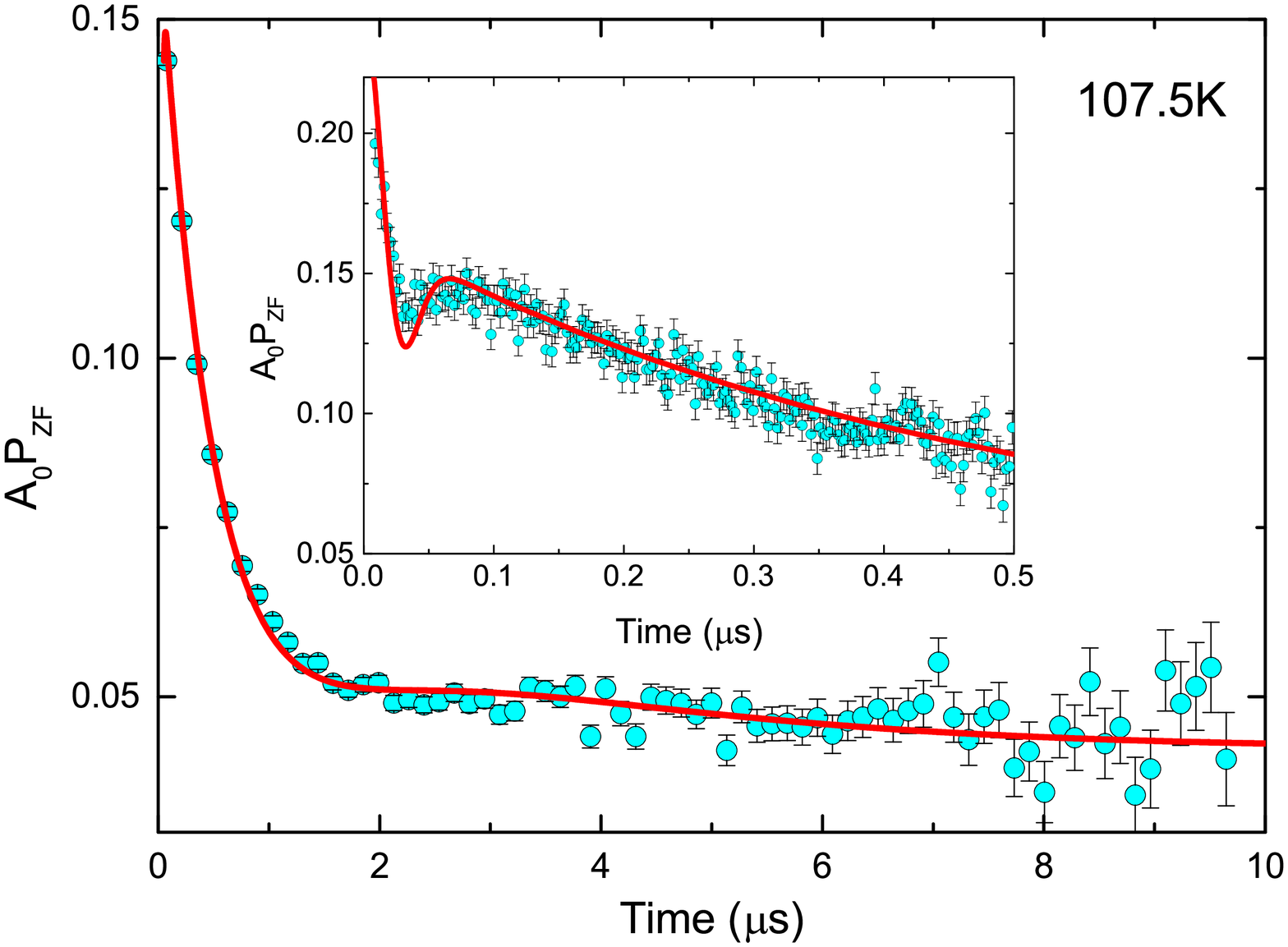}}
\hspace{10mm}
\subfloat[]{\hspace{-8mm}\label{fig:6b}\includegraphics[width=75mm,height=57.5mm]{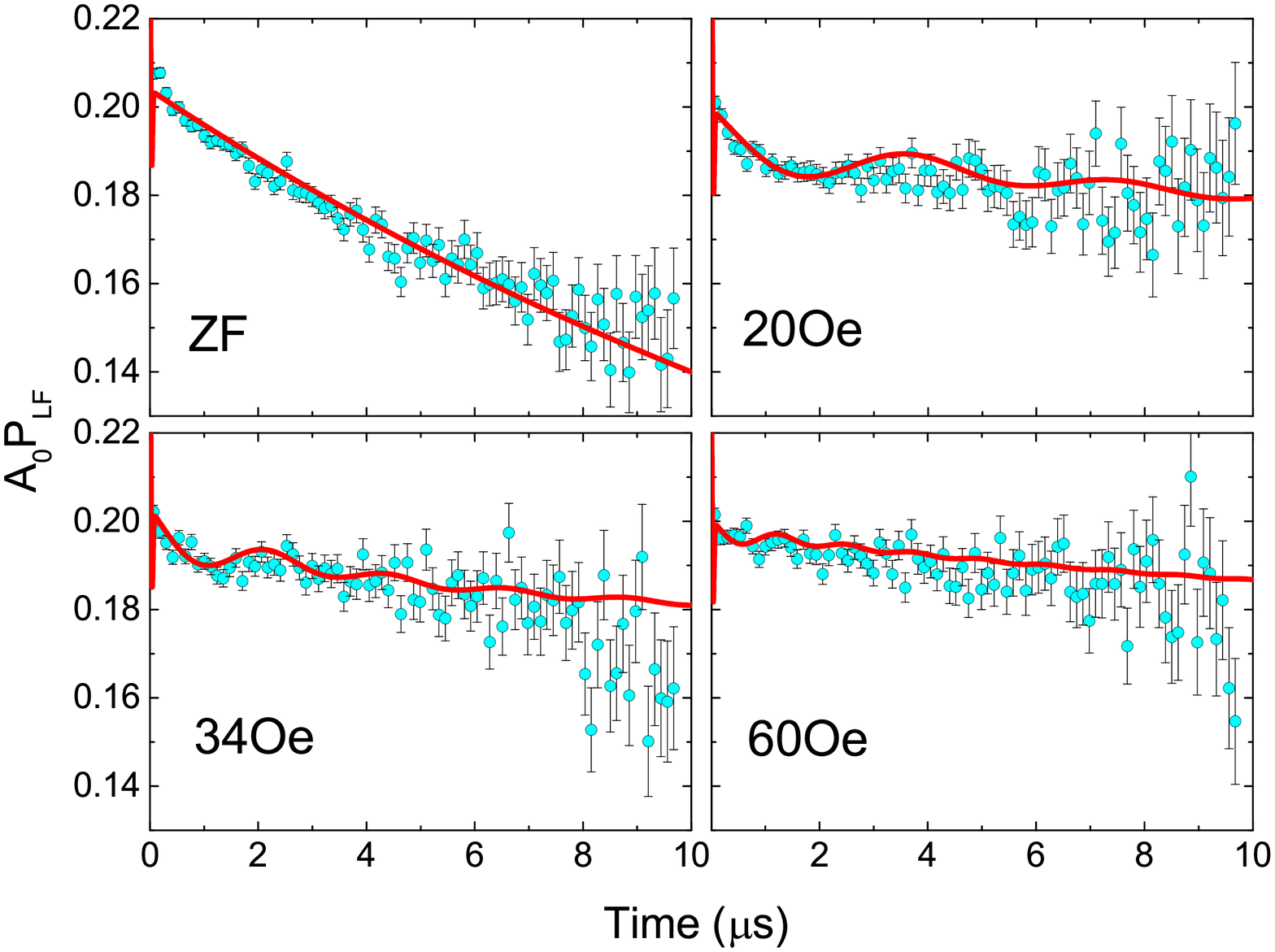}}
  \end{center}
  \vspace{0.5mm}
  \caption{(a) Zero field time spectra at $T=107.5$~K fitted with Eq.(\ref{eq:ZFh}). The inset shows the depolarization function at early times. (b) The zero and longitudinal field depolarization spectra for ${\rm LF}=20, 34, 60$~Oe, at $T=175$~K, also fitted with Eq.(\ref{eq:ZFh}).}
  \label{fig:Fig6}
\end{figure*}
Above the transition, the spectrum consists of a rapidly and slowly relaxing part as shown in Fig.\ref{fig:6a}, in the short and long time domain of the zero field time spectra. The fast relaxing part is related to magnetic domains still present in the sample, revealed also by the $M(H)$ measurement. The slow relaxing part describes a paramagnetic state enriched with dilute magnetic moments. The amplitudes $A_{\rm G}$ and $A_{\rm L}$ are a measure of the volume fractions of the magnetic and paramagnetic regions respectively. With increasing temperature the magnetic regions are diminished and the paramagnetic state is eventually prevalent [Figs.\ref{fig:4a}, \ref{fig:2b}].\par
In Figs.\ref{fig:5a} and \ref{fig:5b}, we present the field distribution width ($\Delta_{\rm GKT}$ and $\Delta_{\rm LKT}$) and field fluctuation rate ($\nu_{\rm GKT}$ and $\nu_{\rm LKT}$) of the Gaussian and Lorentzian KT functions respectively. An increase in dynamics appears, as expected, close to the transition temperature. It is noteworthy that $\Delta_{\rm GKT}$ exhibits a new local maximum at $T_f=117$~K, a phenomenon similar to the one observed in the sister compound LaSrNiReO$_6$ \cite{forslund2020intertwined}. Possibly this maximum indicates the presence of another phase transition that is invisible to bulk magnetization measurements. A dynamic glass-like crossover through local freezing of the components of a random dense magnetic state described by the Gaussian distribution, which in this case is too broad for any oscillatory signal to appear.\par
\begin{figure}[htbp]
  \begin{center}
\subfloat{\hspace{1mm}\includegraphics[width=81mm,height=56mm]{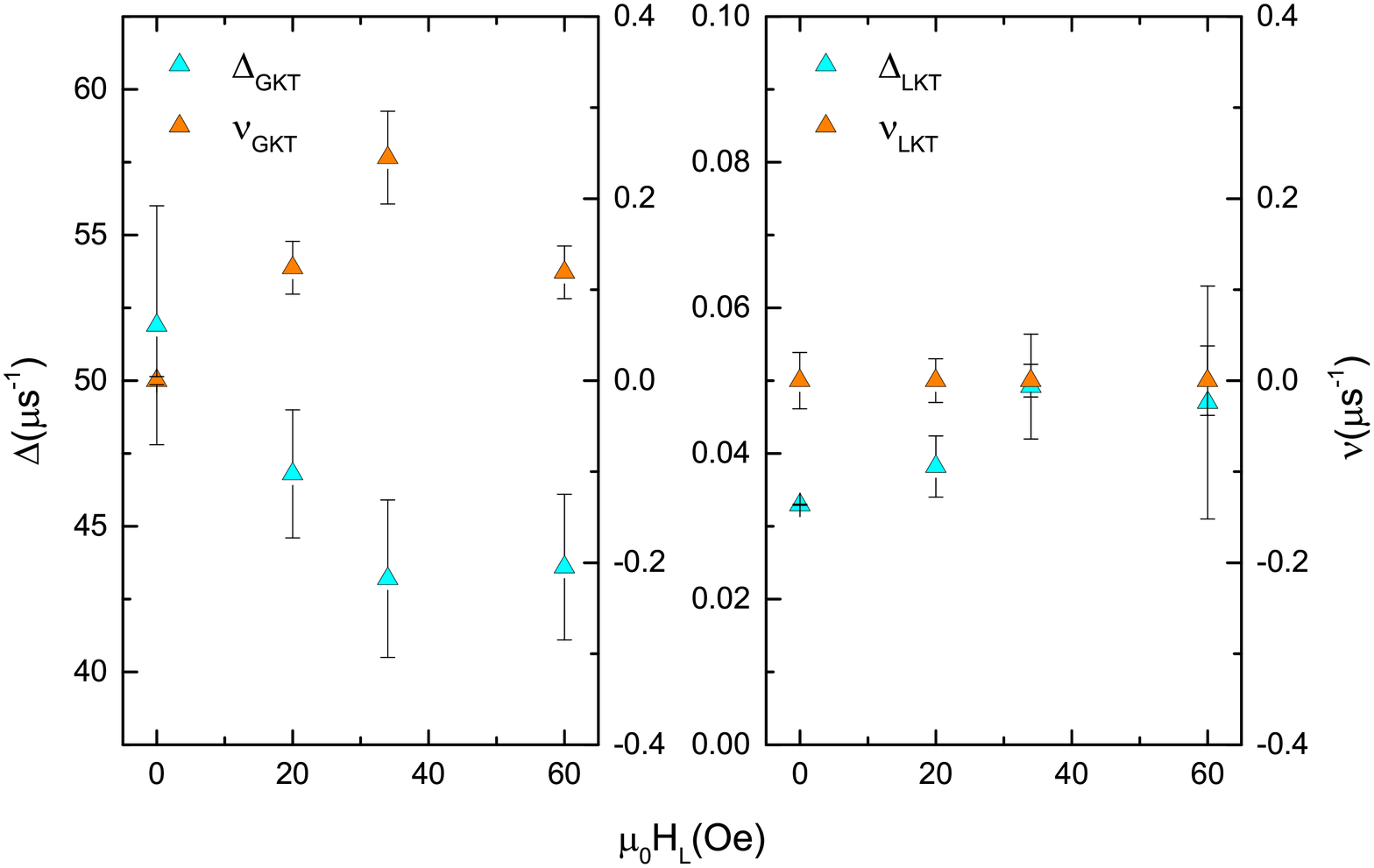}}\quad
  \end{center}
  \vspace{1mm}
  \caption{Field distribution width and depolarization rate for the Gaussian and Lorentzian KT of Eq.\ref{eq:ZFh} as a function of the applied longitudinal field. }
  \label{fig:7}
\end{figure}

This complex magnetic arrangement was further examined by measuring the longitudinal field dependence of the $\mu^+$SR spectra at $175$~K. The ZF- and LF-$\mu^+$SR spectra under fields of $20$, $34$ and $60$~Oe are presented in Fig.\ref{fig:6b}. The LF-$\mu^+$SR spectra are found to be successfully fitted with the same Eq.(\ref{eq:ZFh}) as for the ZF-$\mu^+$SR spectra. Focusing on the slow end we observe a clear decoupling behaviour, an indication of a static but random field distribution in this domain. Indeed the lorentzian KT parameters presented in Fig.\ref{fig:7} show an arguably static field distribution. If we now turn our focus on the fast end of the spectra, the fast relaxing component persists. Overall, a large distribution width and its fluctuation rate ($\Delta_{\rm GKT}$ and $\nu_{\rm GKT}$) suggest that $25\%$ of the volume is characterized by electronic moments in a dynamic state. This is roughly consistent with the wTF-$\mu^+$SR result, in which about $20\%$ of the sample is in a magnetic state at temperatures between 150 and 250K [see Fig.\ref{fig:2b}].
\section{\label{sec:discussion}Discussion}
Both conventional bulk susceptibility and microscopic $\mu^+$SR experiments were used to complete the picture of the magnetically diverse double perovskite LaCaNiReO$_{6}$ down to a long-range ferrimagnetic order established below $T_{\rm N}=103$~K. The underlying superexchange interactions between first, second or higher order neighbours define the magnetic ground state. Alterations to the electron configuration, as a result of spin orbit coupling, the Jahn-Teller effect as well as the composition of the crystal lattice, will result to a frustrate structure. In our case, exchange for a relatively smaller alkaline earth metal in our sample results to a different ground state realised in its sister compound LaSrNiReO$_{6}$, for which an incommensurate magnetic ground state was suggested \cite{forslund2020intertwined}. These two compounds are a characteristic example proving that frustration together with superexchange interaction play a definitive role in the magnetic ordering of perovskite oxides.\par

As GKA rules predict, for a $~152^{\circ}$ Ni-O-Re angle, there exists an antiferromagnetic (AF) exchange interaction between Ni$^{2+}$ and Re$^{5+}$ ions, a signature of geometrical frustration. The long range Ni-Ni and Re-Re interactions in the two sublattices are also significant, as they were described in a similar case for LaSrNiRuO$_{6}$ \cite{ou2018magnetic}. The NNN couplings are expected to be of ferromagnetic nature, leading to ferromagnetically ordered sublattices that are antiferromagnetically coupled. In parallel, the strong spin-orbit coupling of Re atoms, in comparison to Ni ones, reduces the total magnetic moment of the former but also facilitates a DM exchange interaction and antisymmetric superexchange coupling \cite{thio1988antisymmetric}, leaving open the possibility to observe frozen spin components perpendicular to the ferrimagnetic order \cite{ryan2004zero,ryan2000muon}.\par
In our experiments we followed the evolution of magnetic phases with temperature as sublattices order and consequently AF bonds are introduced. Based on the ZF-$\mu^+$SR result [Figs.\ref{fig:3a} and \ref{fig:3c}], we conclude that magnetic domains appear at $T_{\rm N}$. The depolarization lineshape consists of a highly damped oscillation and a slowly relaxing part which correspond to the static magnetic domains and the dynamically arranged magnetic fraction respectively. The TF measurements [Figs.\ref{fig:2a} and \ref{fig:2b}] revealed that the full asymmetry is not recovered until above $T=270$~K where the sample enters the paramagnetic state. In the $103~{\rm K}<T<230~{\rm K}$ region a mixed magnetic phase appears. This evolution is corroborated by the inverse dc susceptibility [Fig.\ref{fig:1b}] behaviour until we reach this high temperature region described by the Curie-Weiss law. Both results hint towards non-single phase sample although the $P2_1/n$ space group fits the NPD patterns at 300, 125 and 1K. De Teresa \textit{et al} in the study of the magnetically similar Ca$_{2}$FeReO$_{6}$ suggested a mesoscopic phase separation in two monoclinic phases that coexist in a wide temperature range, with different magnetic properties \cite{Teresa2004}. This picture is compatible with our results above and below $T_{\rm N}$ and the appearance of two phases can be justified as a result of magnetostriction due to changes in the distance and angle of the B-O-B' bonds.\par 
We now focus on the magnetic ordering below $T_{\rm N}$. In the recorded hysteresis loop at 5~K, shoulders appear at remanent magnetisation. The origin of this phenomenon is thought to be an ordering of transverse spin components due to DM interaction. The monoclinic structure facilitates DM exchange coupling below the ferrimagnetic spin ordering at $T_{\rm N}$ \cite{nguyen1995magnetic,goodenough1994peculiar}. The strength of this interaction is proportional to the spin orbit coupling constant which will be significant for the heavy transition metal, such as, Re. This interplay may result to a portion of the spin components ordering perpendicular to the ferrimagnetic order, as predicted for magnetically frustrated systems \cite{ryan2004zero}.\par
To substantiate this claim we have used the ZF-$\mu^{+}$SR technique which is capable to unambiguously separate static from dynamic signatures compared to bulk magnetization measurements. Below $T_{\rm N}$ the Ni and Re sublattices order \cite{jana2019revisiting,wiebe2002spin,wang2010first} and the critical exponents extracted from this transition follow the mean field model. For a ferrimagnet, this model consists of two alternating sublattices with unequal and antiparallel magnetic moments. The molecular field consists of three coefficients, two ferromagnetic for each sublattice and one antiferromagnetic for their interaction. In the $30~{\rm K}<T<100~{\rm K}$ region the spin lattice is relaxing towards a static commensurate order. The ZF-$\mu^+$SR depolarization data can be understood by a spin wave treatment below $T_{\rm N}$, where both inter- and intra-sublattice exchange is considered, resulting in a two magnon process [Figs.\ref{fig:4b}]. An internal field function was used to describe the ZF muon depolarization spectra and an oscillation is clearly observed up to $0.10~\mu s$ at base temperature $T=2$~K. The corresponding frequency and asymmetry components display a transition to a commensurate magnetic order. This result is in agreement with the DC and AC susceptibility measurements. The dynamic relaxation rates typically display a maximum as the dynamics increase around transition and decay as the dynamics slow down with decreasing temperature. However both $\lambda_{\rm L}$ and $\lambda_{\rm T}$ exhibit a peculiar behaviour between $30~{\rm K}<T<100~{\rm K}$. This may be another indicator for a dynamic phase transition as discussed above, a signature of canting of spin components \cite{ryan2004zero}. In addition to this evolution of the depolarization rate components, we observe a transition of the TF asymmetry in two stages between 50 and 130K, and the appearance of a frequency dependent peak at 50K in the AC susceptibility. These are characteristic features of a spin canted magnetic structure at low temperatures, originating from DM antisymmetric superexchange \cite{moriya1960new,dzyaloshinsky1958thermodynamic}. In that case the spins of the two sublattices will be correlated through a canting angle $\theta$ that will give the ratio of the total to sublattice magnetization \cite{thio1988antisymmetric,bonesteel1993theory}.\par
Above $T_{\rm N}$, the ZF- and LF-$\mu^+$SR spectra are described by a dynamic and a static field distribution function. The fast relaxing component persists at increasing temperature and longitudinal field. We interpret this behaviour as the footprint of a dynamic, random, dense magnetic state. A peak in the distribution width [Figs.\ref{fig:5a}] may signify that spin clusters locally freeze out in this phase. The slow relaxing component prevails as temperature increases. A static field distribution is expected in this domain. The origin of this distribution is most likely a nuclear magnetic field created by the nuclear magnetic moments of the isotopes. A nuclear magnetic field is typically described by a Gaussian distribution, however in our case a GKT function or a combination of distributions did not produce an acceptable fit. A LKT was used instead, assuming randomly distributed nuclear magnetic moments as well as some very dilute Ni and/or Re electronic moments.
\section{\label{sec:conclusion}Conclusions}
We have utilised magnetometry and muon spin spectroscopy to elucidate the magnetic properties of the double perovskite compound LaCaNiReO$_{6}$. As the Ni and Re sublattices mutually order, magnetic phases appear as early as $T<230$~K. With decreasing temperature these phases evolve and finally a transition into a commensurate ferrimagnetic order occurs below $T_{\rm N}=103$~K. As a result of geometrical frustration of the crystal structure, we also find combined microscopic and bulk evidence of a dynamic phase of spin arrangements for $30~{\rm K}<T<100~{\rm K}$. A canting of spins, which does not compromise the static ferrimagnetic order down to base temperature at $2$~K, is a probable scenario. It is a suggestive observation, that although both LaCa$_{x}$Sr$_{1-x}$NiReO$_{6}$, $x=1,0$ share a common dense and dilute magnetic phase above their magnetic order transition and up to $230$~K, the substitution of a larger with a smaller diameter alkaline-earth drastically facilitates or hinders the formation of magnetic order at low temperatures.

\begin{acknowledgments}
We thank Dr. J.-C. Orain for experimental support. This research was supported by the European Commission through a Marie Sk{\l}odowska-Curie Action and the Swedish Research Council - VR (Dnr. 2014-6426 and 2016-06955) as well as the Carl Tryggers Foundation for Scientific Research (CTS-18:272). J.S. acknowledge support from Japan Society for the Promotion Science (JSPS) KAKENHI Grant Nos. JP18H01863 and JP20K21149. Y.S. is funded by the Swedish Research Council (VR) through a Starting Grant (Dnr. 2017-05078) and E.N. the Swedish Foundation for Strategic Research (SSF) within the Swedish national graduate school in neutron scattering (SwedNess). Y.S. and K.P. acknowledge funding a funding from the Area of Advance- Material Sciences from Chalmers University of Technology. D.A. acknowledges partial financial support from the Romanian UEFISCDI Project No. PN-III-P4-ID-PCCF-2016-0112. GS is supported through funding  from the European Union’s Horizon 2020 research and innovation programme under the Marie Sklodowska-Curie grant agreement No 884104 (PSI-FELLOW-III-3i). F.O.L.J acknowledges support from the Swedish Research Counsil - VR (Dnr. 2020-06409). The MPMS used to perform the measurements at the LMX laboratory of the Paul Scherrer Intitute was supported by the Swiss National Science Foundation through grant no. 206021-139082. All images involving crystal structure were made with the VESTA software \cite{Vesta}.
\end{acknowledgments}
\bibliography{Refs} 
\end{document}